\documentclass[aps,prresearch,subeqns.floatfix,amsmath,twocolumn,longbibliography]{revtex4-1}
\usepackage[pdftex]{graphicx}
\usepackage{amsmath}
\usepackage{amsfonts}
\usepackage{graphicx}
\usepackage{hyperref}
\usepackage{comment}
\usepackage{mathtools}
\usepackage{verbatim}
\usepackage[outdir=./]{epstopdf}
\usepackage{subfigure}
\usepackage{hyperref}
\usepackage{latexsym}
\usepackage{dcolumn}
\usepackage{epsf}
\usepackage{float}

\usepackage{color}
\usepackage[utf8]{inputenc}
\begin{document}
\title{
Generalized Adaptation-Induced Non-universal Synchronization Transitions in Random Hypergraphs
}
\author{Sangita Dutta$^1$}
\email{sangitaduttaprl@gmail.com}

\author{Pinaki Pal$^1$}
\email{ppal.maths@nitdgp.ac.in}
\author{Chittaranjan Hens$^2$}
\affiliation{$^1$Department of Mathematics, National Institute of Technology, Durgapur~713209, India}
\affiliation{$^2$Center for Computational Natural Science and Bioinformatics, International Institute of Informational Technology, Gachibowli, Hyderabad 500032, India}

\begin{abstract}
We investigate the effect of partial order parameter adaptation {in form of general functions} on the synchronization behavior of coupled Kuramoto oscillators on top of random hypergraph models. The interactions between the oscillators are considered as pairwise and triangular. Using the Ott-Antonsen ansatz, we obtain a set of self-consistent equations of the order parameter that describe the synchronization diagrams.  
A broad diversity of synchronization transitions are observed as a result of the interaction between the partial adaptation approach, generalized adaptation functions, and coupling strengths. The system specifically shows a double-jump transition under a power-law form of the adaptation function. A polynomial form of the adaptation function leads to the emergence of an intermediate synchronization state for specific combinations of one negative and one positive coefficient. Moreover, the synchronization transition may become continuous or explosive when the pairwise coupling strength varies.
 The generality of this synchronization behavior is further supported by results obtained using a Gaussian adaptation function.

\end{abstract}
\maketitle
\section{Introduction}
One of the most fascinating collective behaviors of complex systems is synchronization, resulting from nonlinear interactions between the units. In a variety of systems, the Kuramoto model is essential for capturing this fascinating phenomenon \cite{kuramoto1984chemical,strogatz2000kuramoto,Pikovsky_synchronization_book,rodrigues2016kuramoto}. In view of its significance in real-world systems, researchers focus a lot of attention on investigating the paths to synchronization and desynchronization in networked systems \cite{boccaletti2006complex}. Under a unimodal frequency distribution, the classical Kuramoto model has been found to exhibit a continuous transition to synchronization \cite{rodrigues2016kuramoto,dorfler2011critical}, which is typified by a progressive rise in the degree of synchronization following a critical coupling strength.
The magnitude of the Kuramoto order parameter indicates the degree of synchronization. Explosive synchronization transition may also result from the interaction of various network topologies, frequency distributions, and coupling configurations, such as degree-degree correlation \cite{gomez2011explosive}, degree-frequency correlation \cite{coutinho2013kuramoto,kundu2017transition,kundu2019synchronization}, uniform and bimodal frequency \cite{pazo2005thermodynamic,pazo2009existence}, adaptive coupling \cite{filatrella2007generalized, khanra2020amplification, khanra_chaossoli2021, Zhang_PRL2015}, multiplexing \cite{khanra2018explosive}, etc. It is distinguished by an abrupt shift from an incoherent to a synchronized condition. 
\par


Depending on the phase-lag, coupling functions, frequency distributions and network structure, the system may also show  a nontrivial route to synchronization transition \cite{omel2012nonuniversal,omel2013bifurcations,omel2016there}. For instance, it may follow a weak synchronization path before showing a sharp jump in the order parameter \cite{skardal2022tiered,rajwani2023tiered,dutta2024transition}. This is called tiered synchronization. 
According to current research, the frequency distribution has an impact on multistep synchronization transitions in adaptively coupled systems \cite{fialkowski2023heterogeneous}. Hypergraphs with higher-order coupling under power-law adaptation have most recently been found to undergo a two-step transition from incoherence to partial coherence, followed by a jump to stronger synchronization \cite{dutta2025double}. This model is notable for its analytical tractability, which enables adjustment of parameters like coupling strength, adaptation exponent, and fraction of adapted nodes to control transitions path, create double explosive transition, and tune hysteresis width. Moreover, it has been shown in \cite{costa2025exact} that specific choice of higher-order coupling strengths, considered upto fifth order interactions may induce double jumps in the synchronization transitions.
\par
It is important to note that an explosive synchronization transition is naturally induced and promoted by the presence of higher-order interactions in networked systems, independent of any correlation between the dynamics and structure of networks \cite{skardal2020higher,adhikari2023synchronization,dutta2023impact,dutta2023perfect,smith2024determining}. In addition to bistability, these higher-order interactions can also cause multistability \cite{skardal2019abrupt}, which is the coexistence of several stable states in a parameter regime. An excellent illustration of this is tiered synchronization, in which states of incoherence, weak synchronization, and strong synchronization coexist. A time delay \cite{skardal2022tiered} in the extended Kuramoto model or a change in the order parameter in the Kuramoto or Sakaguchi-Kuramoto model \cite{rajwani2023tiered,dutta2024transition, manoranjani2023phase,biswas2024effect} can be used to achieve this kind of transition. However, all these methods did not reveal a double explosive transition in random hypergraphs.
\par Motivated by recent findings on double explosive synchronization transitions \cite{dutta2025double}, we want to investigate the coupling function's impact on synchronization (continuous, explosive, double explosive, or coexisting multistable states) dynamics in greater detail. We specifically concentrate on adaptive coupling strategies in which the coupling strength is determined by nonlinear or non-monotonic functions that depend on the global order parameter. 
Inspired by power-law functions \cite{Zhang_PRL2015,xu2021collective}, polynomial functions have been used as adaptation mechanisms \cite{manoranjani2023phase}. A variety of physical systems, including Josephson junctions, have been modeled using these nonlinear forms \cite{filatrella2007generalized}. Gaussian functions have also been proposed to represent more complicated, non-monotonic adaptation behaviors \cite{jin2023synchronization}. 
\par Thus, it is reasonable to wonder what would happen if such general functions were adapted and applied selectively to a subset of nodes in a higher-order interaction environment. Remember that the transition routes (as a function of pairwise coupling) were evaluated from explosive to continuous paths via double jumps involving two hysteresis under the power law adaptation function.
In this work, we investigate synchronization transitions by using the coupling strengths to partially adjust generic functions of the order parameter. As in \cite{dutta2025double}, we employ a self-consistent approach to construct a pair of analytical equations, the solutions of which yield a variety of nontrivial synchronization diagrams. Numerical simulations support these analytical predictions well. Two important findings are presented. First, the system shows a doubly explosive transition when adjusting the higher-order coupling strength $K_2$ under partial adaptation via a power-law function of the order parameter. 
Second, for polynomial adaptation functions, the transition routes are strongly influenced by the signs and magnitudes of the coefficients. Either an explosive or continuous transition is seen in the system when both coefficients are positive. On the other hand, multistability—coexisting incoherent, weakly synchronized, and strongly synchronized states—occurs when one coefficient is positive and the other is negative. In this regime, a nontrivial intermediate state is revealed as the degree of weak synchronization diminishes as the higher-order coupling strength increases.
These results provide a deeper understanding of how adaptive processes influence the complexity of synchronization transitions, particularly when higher-order interactions are present.

\section{Model Description}
The evolution of phases of coupled oscillators can be governed by the equations proposed by Kuramoto, given as
\begin{eqnarray}
\label{model_net}
\dot{\theta_i}&=&\omega_i+K_1 g_1(r_1) \sum_{j=1}^{N} A_{ij} \sin(\theta_j-\theta_i) \nonumber\\
&+&K_2 g_2(r_1)\sum_{j=1}^N \sum_{k=1}^N B_{ijk} \sin(2\theta_j-\theta_k-\theta_i), \\ &&\hspace*{0.6cm} i=1,2,\dots,N, \nonumber
\end{eqnarray}
where $\theta_i$ is the phase of the $i$th oscillator, $\omega_i$ is the $i$th component of the natural frequency $\omega$, distributed according to a distribution function $G(\omega)$. $K_1$ and $K_2$ denote the pairwise and triadic coupling strength, respectively. $A_{ij}$ is the $ij$th entry in the adjacency matrix of the considered network that represents the pairwise connections between nodes, while $B_{ijk}$ represents the triangular connections. $N$ is the total number of nodes in the system.
To quantify the level of synchronization of the networked system, the local order parameters are defined by
\begin{eqnarray}
    R_i^1=\sum_{j=1}^{N}A_{ij}e^{i\theta_j}, \hspace{0.2cm} R_i^2=\sum_{j,k=1}^{N}B_{ijk}e^{2i\theta_j}e^{-i\theta_k} 
    \label{local_order_parameter}
\end{eqnarray}
and the global order parameters are defined by
\begin{eqnarray}
     z_1=r_1(t)e^{i\psi_1}&=&\frac{1}{N\langle k^{(1)}\rangle}\sum_{i=1}^N R_i^1, \nonumber\\ z_2=r_2(t)e^{i\psi_2}&=&\frac{1}{2N\langle k^{(2)}\rangle}\sum_{i=1}^N R_i^2,
     \label{global_order_parameter}
\end{eqnarray}
where $r_1$ and $r_2$ are the magnitudes, $\psi_1$ and $\psi_2$ are the average phase values. $\langle k^{(1)}\rangle$ and $\langle k^{(2)}\rangle$ denote the mean pairwise and triadic degrees, respectively. In this study, we will concentrate on the behavior of the global order parameter $r_1$ under adaptation technique with pairwise and higher-order coupling strengths in the form of two functions $g_1(r_1)$ and $g_2(r_1)$, respectively. We have considered mainly two adaptation functions, one is polynomial function and another is gaussian function. To accomplish this, we have constructed a synthetic random hypergraph where the nodes are joined by links with probability $p^{(1)}=\frac{\langle k^{(1)} \rangle }{N}$, which is a classic Erd{\H{o}}s-R{\'e}nyi (ER) network and the nodes are joined by triangles with probability $p^{(2)}=\frac{2\langle k^{(2)} \rangle }{N^2}$. We have denoted $\langle k^{(1)}\rangle=\langle k \rangle$ and $\langle k^{(2)}\rangle=\langle q \rangle$. In the next section, we move to obtain a low-dimensional model to analyze the system easily.
\section{Derivation of Low-Dimensional Model}
At first we have rewritten Eq.(\ref{model_net}) in vector form as  
\begin{equation}
\label{reduction_model_net}
\dot{\theta_i}=\omega_i+\frac{1}{2i}\left[e^{-i\theta_i}H_i-e^{i\theta_i}\bar{H_i}\right],
\end{equation}
where $H_i=K_1 g_1(r_1) R_i^1+K_2 g_2(r_1) R_i^2$. At this point it is convenient to assume the nodes with same hyperdegree are equivalent. Therefore, the local order parameters can be expressed as 
\begin{eqnarray*}
&&R_i^1\rightarrow R^1(k_i,t), \\
&&R_i^2\rightarrow R^2(k_i,t).
\end{eqnarray*}
Following the literature on Kuramoto oscillators, we move to the continuum description of the considered system by letting $N \rightarrow \infty$. Consequently, two density functions can be introduced. one is $f(\theta, \omega, k, t)$ describes the density of oscillators with phase $\theta$, hyperdegree $k$ and natural frequency $\omega$ at time $t$ and another is $F(\theta',\omega',\theta'',\omega'',k',k'',t)$ describes the joint density of two oscillators with phase, frequency, and hyperdegree of $\theta',~\omega',~k'$ and $\theta'',~\omega'',~k''$. Now following Ref.\cite{adhikari2023synchronization}, we made the assumption that the joint density function $F(\theta',\omega',\theta'',\omega'',k',k'',t)$ can be decomposed into the product of two density functions in the limit of complete incoherent and synchronized states or in the case of dense hypergraphs. Therefore, we can write
\begin{equation*}
  F(\theta',\omega',\theta'',\omega'',k',k'',t)=f(\theta',\omega',k',t) f(\theta'',\omega'',k'',t).
\end{equation*}
Furthermore, in the continuum limit the order parameters $R^1(k)$ and $R^2(k)$ can be expressed as  
  \begin{eqnarray}
    R^1(k)&=&\sum_{k'}N(k')p^{(2)}\int\int f(\theta',\omega',k',t)e^{i\theta'}d\theta' d\omega', \label{eq_R1}
    \end{eqnarray}
    {\scriptsize{
    \begin{eqnarray}
    R^2(k)&=&\sum_{k',k''}N(k')N(k'')p^{(3)} \\
    &&\int\int\int\int F(\theta',\omega',\theta'',\omega'',k',k'',t)e^{2i\theta'}e^{-i\theta''}d\theta' d\omega' d\theta'' d\omega'' \nonumber \\
    &\approx& \sum_{k',k''}N(k')N(k'')p^{(3)}(k,k',k'')\int\int f(\theta',\omega',k',t) e^{2i\theta'}d\theta' d\omega' \nonumber \\
   && \int\int f(\theta'',\omega'',k'',t) e^{-i\theta''} d\theta'' d\omega'',    \label{eq_R2}
\end{eqnarray}  }}
where $N(k)$ is the number of nodes having degree $k$.
Due to the conservation of the oscillators in the system, the density function $f$ must satisfy the continuity equation, given by
\begin{equation}
    \frac{\partial f}{\partial t}+\frac{\partial}{\partial \theta}{\left(f\left(\omega+\frac{1}{2i}\left[e^{-i\theta}H-e^{i\theta}\bar{H}\right]\right)\right)}=0.
    \label{continuity_eq2_supp}
\end{equation}
Because of fixed natural frequency of the oscillators and periodic nature of the density function, we can expand the density function in the Fourier series as
\begin{eqnarray}
    f(\theta,\omega,k,t)=\frac{G(\omega)}{2\pi}\left[1+\sum_{n=1}^\infty \left[f_n e^{in\theta}+\bar{f}_ne^{-in\theta}\right]\right]. 
    \label{f_fourier}
\end{eqnarray}
Where, $f_n$ is the $n$th coefficient and $\bar{f}_n$ is the complex conjugate of $f_n$. Then, in order to analyze Eq.(\ref{continuity_eq2_supp}) and finding a low dimensional description of it, we follow the Ott--Antonsen ansatz \cite{ott2008low}. It allows us to consider the form of $f_n$ as $f_n=\alpha^n$, with $\alpha$ being analytic function and $|\alpha|\leq 1$. This form of the coefficients of the above series assure its convergence.
In the next step we insert this ansatz in the continuity Eq. (\ref{continuity_eq2_supp}), yields
\begin{equation}
\label{eq_alpha_net}
\dot{\alpha}+i\alpha\omega-\frac{1}{2}\left[\bar{H}-H\alpha^2\right]=0.
\end{equation}
Therefore, the considered system has been reduced to a single differential equation. In order to find the order parameter values we have substituted the Fourier expansion of $f$ into Eqs. (\ref{eq_R1}) and (\ref{eq_R2}) and obtain
    \begin{eqnarray}
        R^1(k)&=&\sum_{k'}N(k')p^{(2)}\int\int\frac{G(\omega')}{2\pi}[1+\sum_{n=1}^\infty [\alpha^n e^{in\theta'} \nonumber\\
        &+&\Bar{\alpha}^ne^{-in\theta'}]]e^{i\theta'}d\theta' d\omega'   \nonumber \\
        &=& \sum_{k'}N(k')p^{(2)}\int G(\omega')\bar{\alpha}(\omega',k',t) d\omega', \label{eq_R1_cal}
    \end{eqnarray}
    Similarly we calculate $R^2(k)$ and get
    \begin{eqnarray}
      R^2(k)&=&\sum_{k',k''}N(k')N(k'')p^{(3)}\int G(\omega')\bar{\alpha}^2(\omega',k',t) d\omega' \nonumber\\
      &&\int G(\omega'')\alpha(\omega'',k'',t) d\omega''. \label{eq_R2_cal}
    \end{eqnarray}
In this study, we have chosen the natural frequencies from a Lorentzian distribution, $G(\omega)=\frac{\Delta}{\pi[\Delta^2+(\omega-\omega_0)^2]}$, where $\Delta$ is the half width and $\omega_0$ is the peak of the distribution. With this choice of $G(\omega)$ we can easily calculate Eq.\ref{eq_R1_cal} and Eq.(\ref{eq_R2_cal}) by performing the contour integration in the lower-half $\omega$ plane, given by    
    \begin{eqnarray}
        R^1(k)=\sum_{k'}N(k')p^{(2)}\bar{\alpha}(\omega_0-i\Delta,k',t), 
        \label{eq_R1_value}
    \end{eqnarray}
    \begin{eqnarray}
        R^2(k)=\sum_{k',k''}N(k')N(k'')p^{(3)}\bar{\alpha}^2(\omega_0-i\Delta,k',t) \nonumber\\
        \alpha(\omega_0 -i\Delta, k'',t). \label{eq_R2_value}
    \end{eqnarray}
After that, evaluating Eq.(\ref{eq_alpha_net}) at $\omega=\omega_0-i\Delta$ and inserting the order parameter values from Eq. (\ref{eq_R1_value}) and Eq.(\ref{eq_R2_value}) we obtain
\begin{eqnarray}
    &&\dot{\alpha}+i\omega_0\alpha+\Delta\alpha-\frac{K_1g_1(r_1)}{2}\sum_{k'}N(k')p^{(2)}[\alpha(k') \nonumber\\
    &&-\bar{\alpha}(k')\alpha^2(k)]-\frac{K_2g_2(r_1)}{2}\sum_{k',k''}N(k')N(k'') p^{(3)} \nonumber\\
    &&[\alpha^2(k')\bar{\alpha}(k'')-\bar{\alpha}^2(k')\alpha(k'')\alpha^2(k)]=0.    \label{reduced_model_supp}
\end{eqnarray}
This is the required reduced order model of the considered networked system. Now we will substitute the probability values ($p^{(2)}$ and $p^{(3)}$) of the considered random network in the above equation, which yields
\begin{eqnarray}
    &&\dot{\alpha}+i\omega_0\alpha+\Delta\alpha-\frac{K_1g_1(r_1)}{2N}\sum_{k'}N(k')\langle k \rangle[\alpha(k')   \nonumber\\
   && -\bar{\alpha}(k')\alpha^2(k)]-\frac{K_2g_2(r_1)}{2N^2}\sum_{k',k''}N(k')N(k'') 2\langle q \rangle \nonumber \\
   &&[\alpha^2(k')\bar{\alpha}(k'')-\bar{\alpha}^2(k')\alpha(k'')\alpha^2(k)]=0.   \label{eq_alpha_er}
\end{eqnarray}

Thereafter, we have introduced two new variables by assuming
\begin{eqnarray}
    U_1=\frac{1}{N}\sum_{k'}N(k')\alpha(k'), \label{U1_er_supp}\\
    U_2=\frac{1}{N}\sum_{k'}N(k')\alpha^2(k').
    \label{U2_er_supp}
\end{eqnarray}
Substitution of these $U_1$ and $U_2$ into Eq. (\ref{eq_alpha_er}) reduces it to 
\begin{eqnarray}
    \dot{\alpha}(k)&+&i\omega_0\alpha(k)+\Delta\alpha(k)-\frac{K_1g_1(r_1) k }{2}[U_1 -\bar{U}_1\alpha^2(k)] \nonumber \\
    &&-K_2g_2(r_1) k[U_2\bar{U}_1-\bar{U}_2U_1\alpha^2(k)]=0.
\end{eqnarray}
For the sake of obtaining stationary rotating solutions of $\alpha$, we have to use the polar forms as $\alpha(k,t)=\alpha e^{i\omega_1 t}$, $U_1(t)=U_1e^{i\omega_1 t}$ and $U_2(t)=U_2e^{2i\omega_1 t}$. Then, taking $\Delta=1$ we have separated the real and imaginary parts of the above equation we get
\begin{eqnarray}
    \alpha-(\frac{K_1}{2}g_1(r_1)\langle k \rangle U_1+K_2g_2(r_1)\langle q\rangle U_1U_2) \nonumber \\
    (1-\alpha^2)=0,
    \label{eq_alpha_real}
\end{eqnarray}
\begin{eqnarray}
    \alpha\omega_1 =-\alpha\omega_0+(\frac{K_1}{2}g_1(r_1)\langle k \rangle U_1 +K_2g_2(r_1)\langle q\rangle U_1U_2) \nonumber \\
    (1+\alpha^2).
\end{eqnarray} 
Clearly Eq.(\ref{eq_alpha_real}) is a quadratic equation. Therefore, we can solve it to find the value of $\alpha$. This $\alpha$ combined with Eqs. (\ref{U1_er_supp}) and (\ref{U2_er_supp}) forms the self-consistent equations given by
\begin{eqnarray}
    U_1=\frac{1}{N}\sum_{k}N(k)\alpha(k,U_1,U_2), \label{sc_U1}\\
    U_2=\frac{1}{N}\sum_{k}N(k)\alpha^2(k,U_1,U_2),  \label{sc_U2}
\end{eqnarray}
where 
\begin{scriptsize}
\begin{eqnarray}
    \alpha=\frac{-1+\sqrt{1+(K_1g_1(r_1)\langle k\rangle U_1+2K_2g_2(r_1)\langle q\rangle U_1U_2)^2}}{(K_1g_1(r_1)\langle k\rangle U_1+2K_2 g_2(r_1)\langle q\rangle U_1U_2)}.
    \label{sc_alpha}
\end{eqnarray}
\end{scriptsize}
Moreover, substituting the probability values $p^{(2)}$ and $p^{(3)}$ in the definition of the order parameter values (Eqs. (\ref{eq_R1_value}) and (\ref{eq_R2_value})) we get the relation between the order parameters and $U_1$, $U_2$, follows
\begin{eqnarray*}
    z_1(t)&=&\frac{1}{N\langle k \rangle}\sum_{k,k'}N(k)N(k')\frac{kk'}{N\langle k \rangle}\bar{\alpha}(\omega_0-i\Delta,k',t) \\
    &=&\bar{U}_1, \\
  z_2(t) &=& \bar{U}_2 U_1.
\end{eqnarray*}
These relations imply $r_1=|U_1|=U_1$ and $r_2=|\bar{U}_2U_1|=U_2U_1$. Now we can find the order parameter values by solving the self-consistent equations (\ref{sc_U1}), (\ref{sc_U2}), which will illustrate the synchronization profiles for different parameter values.

\section{Results}
\begin{figure}[h!]
    \centering
    \includegraphics[width=1\linewidth]{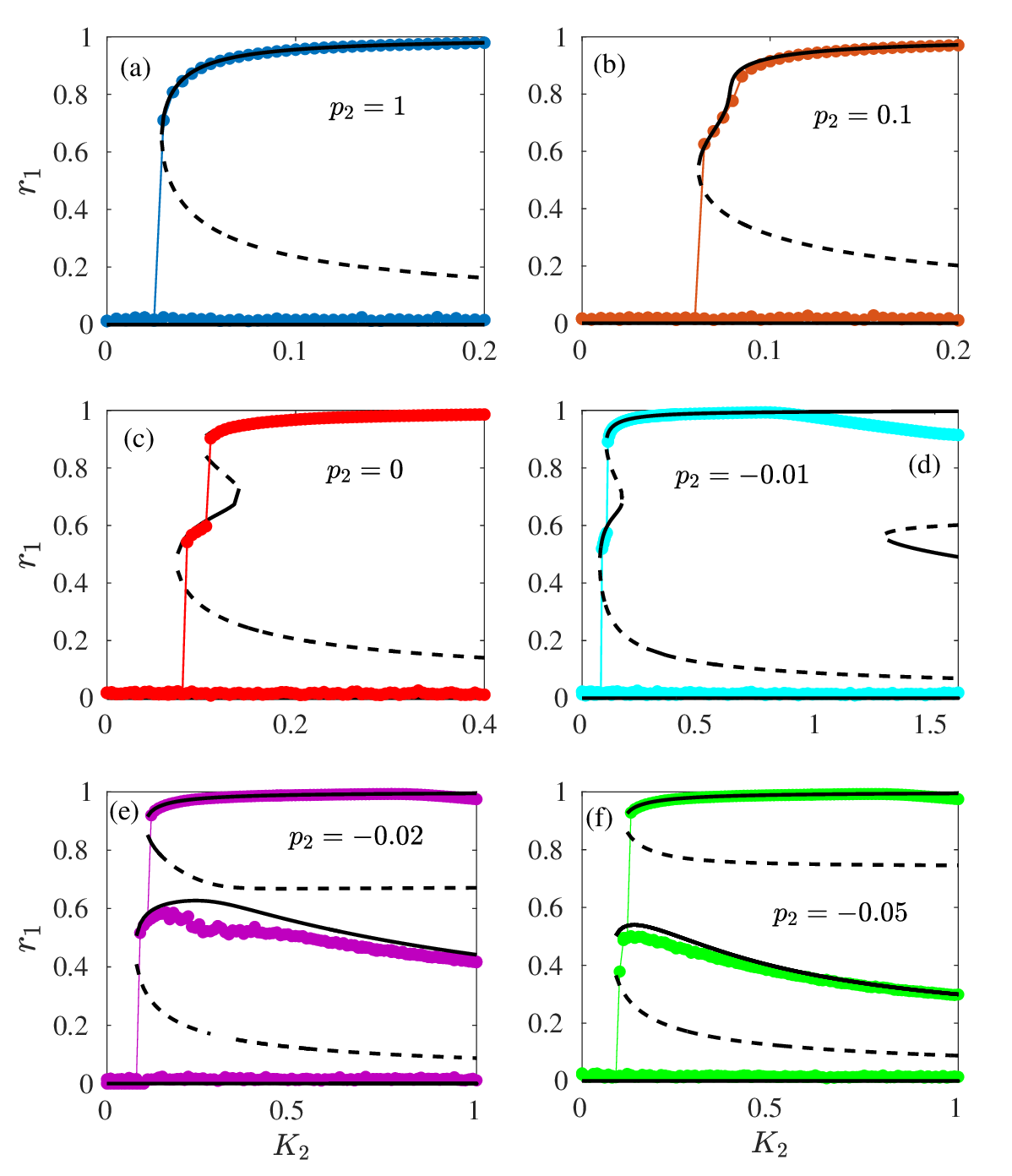}
    \caption{$r_1$ as a function of triadic coupling $K_2$ under the adaptation function $g_{2}=p_1r_1^{b}+p_2$. The adaptation with the triadic coupling is given partially to nodes having degree greater equal to $102$ by taking the coefficient values as (a) $p_2=1$, (b) $p_2=0.1$, (c) $p_2=0$, (d) $p_2=-0.01$, (e) $p_2=-0.02$ and (f) $p_2=-0.05$. Other parameter values are taken as $K_1=0.01,~m_1=1,~m_2=0,~p_1=1,~a=0$ and $b=10$.}
    \label{K2_vs_r1_diff_p2_1}
\end{figure}

In this section, we have investigated the behavior of the synchronization transitions with the variation of triadic coupling. We start our analysis by considering a random network of size $N=5000$ with mean degrees $\langle k^{(1)}\rangle = \langle k^{(2)} \rangle =100$.
First, we take the adaptation functions as polynomials \cite{manoranjani2023phase} of the form $g_1=m_1 r_1^a+m_2$ and $g_2=p_1 r_1^b+p_2$. For simplification, we choose the parameter values as $m_1=1$, $m_2=0$, $p_1=1$, $p_2=0$, $a=0$ and $b=10$. $a=0$ implies no adaptation with the pairwise coupling term. Then we adapted the function $g_2(r_1)$ of global order parameter with the higher-order coupling strength partially to nodes (upto a degree threshold value $d_{th}$) of the considered system. To see the effect of this partial adapting technique on the synchronization profiles, we solved the self-consistent equations by varying the triadic coupling $K_2$ and plot the solutions in Fig. \ref{K2_vs_r1_diff_p2_1}(c). 
Other parameters are kept fixed at $K_1=0.01$ and $d_{th}=102$. We observed a different type of synchronization diagram compared to the classical explosive synchronization transition. Previous works \cite{dutta2023impact,dutta2024transition} reports that the explosive synchronization transition with respect to the triadic coupling is associated with a single saddle node bifurcation with a stable incoherent state. In contrast, here we observed that under the partial order parameter adaptation technique the system undergoes two saddle-node bifurcations along with a stable incoherent state (Fig. 1(c)).

For comparison purpose, We have depicted the evolution of the transition paths from zero adaptation to complete adaptation in Fig.\ref{fig1} for partial control of nodes in presence of power law adaptation ($p_2=0$) (see the appendix).

To verify these analytical findings, we have simulated the system Eq.(\ref{model_net}). The Euler method has been used for numerical integration. After discarding a sufficient amount of transient part we have calculated the global order parameter $r_1$ with the variation of $K_2$. The forward and backward sweeps are carried out by increasing and decreasing $K_2$ in small steps, respectively, to experience hysteresis in the transition path. Initially, the phase of the oscillators are distributed uniformly in $[-\pi,\pi]$. After that, the final phase values of the trajectory corresponding to the previous coupling strength have been used as an initial condition for the next simulation. We take the same parameter values used 
in Fig.\ref{K2_vs_r1_diff_p2_1}(c) and observe that the system jumps twice towards the desynchronization state, following the path described by the solutions of the self-consistent equations. Due to the split in the path, the system is attracted to the generated stable part. Unlike the case of variation of $K_1$ reported in the paper \cite{dutta2025double}, here also this stable part facilitates the desynchronization transition. Since in case of variation $K_2$, the forward transition point does not exist in the thermodynamic limit, therefore, in this case the system shows double explosive jump only in backward simulation. A similar kind of transition has been found numerically in our most recent work \cite{das2025effect}, although the analytical synchronization profiles are quite different from this paper. Also, there the transition paths are dependent on initial conditions. Here, the synchronization transitions are initial condition independent.

\begin{figure}
    \centering
    \includegraphics[width=1\linewidth]{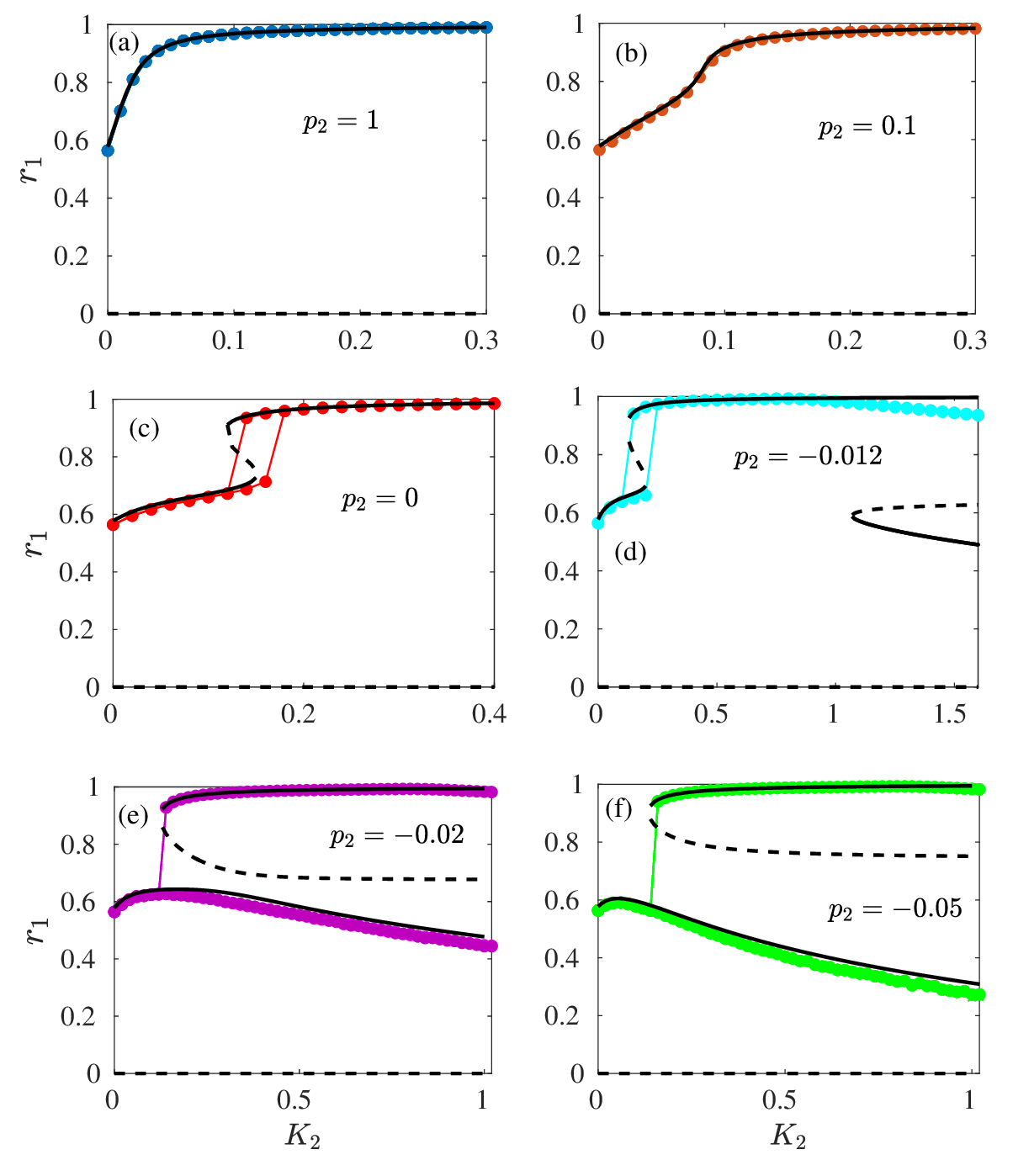}
    \caption{$r_1$ as a function of triadic coupling $K_2$ under the adaptation function $g_{2}=p_1r_1^{b}+p_2$. The order parameter adaptation with the triadic coupling is given partially to nodes having degree greater equal to $90$ by taking the coefficient values as (a) $p_2=1$, (b) $p_2=0.1$, (c) $p_2=0$, (d) $p_2=-0.012$, (e) $p_2=-0.02$ and (f) $p_2=-0.05$. Other parameter values are taken as $K_1=0.03,~m_1=1,~m_2=0,~p_1=1,~a=0$ and $b=10$.}
    \label{K2_vs_r1_K1_0p03_vary_p2_1}
\end{figure}

Now to get better insight of the synchronization diagram under the considered polynomial function we take non zero values of $p_2$ and plot $r_1$ curves by solving the derived self consistent equations. 
We vary the value of $p_2$ from negative to positive while keeping $p_1$ fixed at $1$. 
Figure. \ref{K2_vs_r1_diff_p2_1} demonstrates that as the value of $p_2$ increases, the fold in the synchronization path vanishes and it becomes typical explosive. In addition, the transition point moves backward (towards low coupling strength) with increasing vertical width (distance between stable and unstable state) of the curve $r_1$. The system shows this behavior due to increase in the effective triadic coupling strength. 
Then, we gradually decrease the value of $p_2$.
We have seen that after slight decrease in $p_2$, a new $r_1$ curve pops up in the $K_2-r_1$ space, which has a stable (lower part) and an unstable (upper part) branch. As the value of $p_2$ decreases more, this new curve moves toward the previously existing $r_1$ curve (Fig.\ref{K2_vs_r1_diff_p2_1}(d)) and after a certain value of $p_2$, the new one touches the previous one. Further decrease of $p_2$, split the synchronization path into two disjoint portions. The synchronization level of the upper portion remains strong. However, the synchronization level in the lower portion decreases with increasing value $K_2$. To elucidate these synchronization transitions, we numerically simulate the considered system. It is clear from Fig.\ref{K2_vs_r1_diff_p2_1} that the numerical solutions complement the analytical ones. At $p_2=-0.01$, the transition to desynchronization is the same as in $p_2=0$, under adaptive initial values. While, suitable initial condition will reach the system to the new $r_1$ curve. On the other hand, for $p_2=-0.02$, the forward simulation from the weak synchronization state follows the intermediate stable state with decreasing $r_1$. Similar things happen for $p_2=-0.05$.


\begin{figure}
    \centering
    \includegraphics[width=1\linewidth]{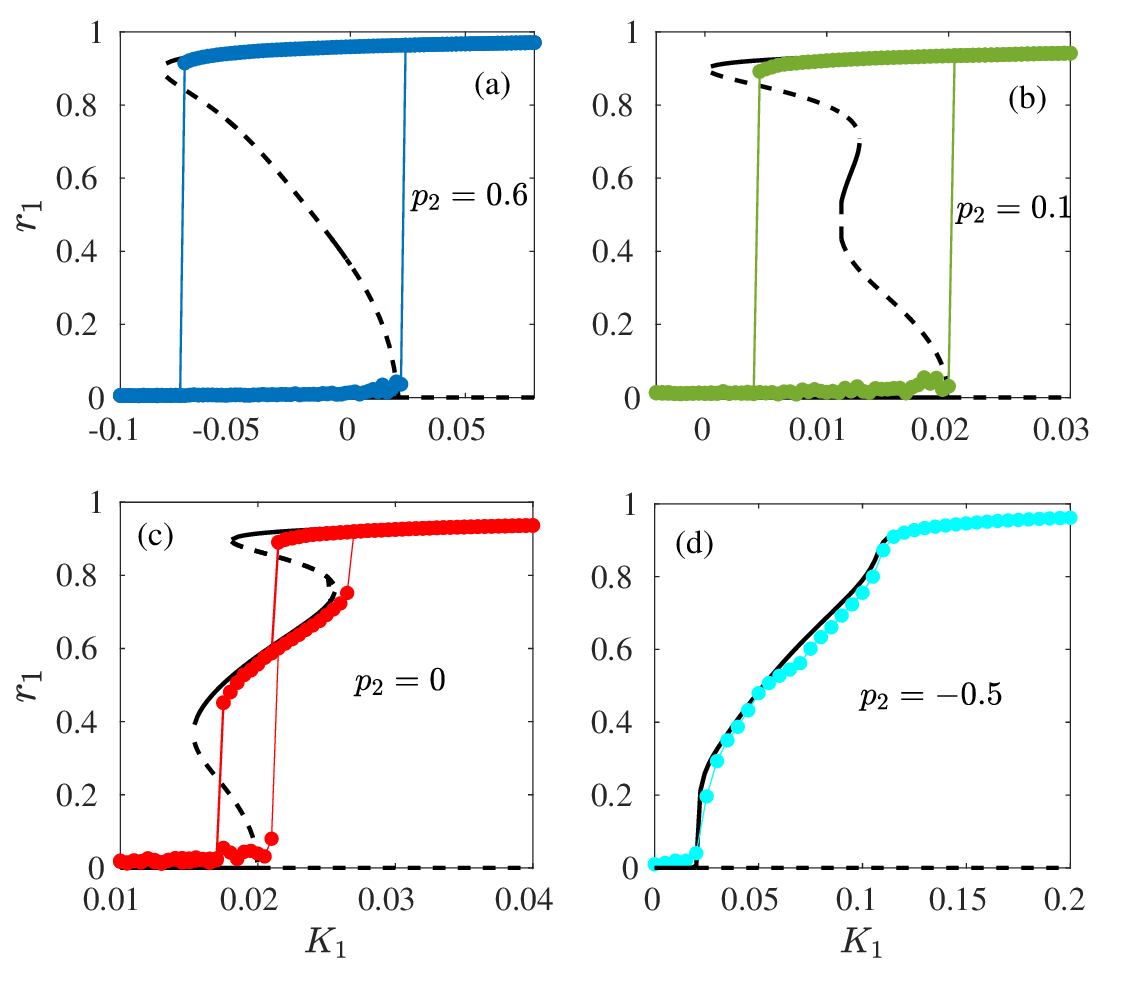}
    \caption{$r_1$ as a function of pairwise coupling $K_1$ under under the adaptation function $g_{2}=p_1r_1^{b}+p_2$. The order parameter adaptation with the triadic coupling is given partially to nodes having degree greater equal to $95$ by taking the coefficient values as (a) $p_2=0.6$, (b) $p_2=0.1$, (c) $p_2=0$ and (d) $p_2=-0.5$. Other parameter values are taken as $K_2=0.12,~m_1=1,~m_2=0,~p_1=1,~a=0$ and $b=10$.}
    \label{K1_vs_r1_adapt_K2_polynomial}
\end{figure}
Note that in this case, the transition remains explosive for high $p_2$ due to low pairwise coupling.
 Then we increase the pairwise coupling strength to $K_1=0.03$ and proceed in a similar manner with other parameters fixed at the same values. We have presented the evolution of transitions from zero adaptation to complete adaptation in Fig. \ref{fig2} (see the appendix for comparison) and have seen that the transition becomes tiered from continuous.
To study the impact of $p_2$, for high $K_1$, we took Fig. \ref{fig2} (b) and varied $p_2$ in a similar way. 
\par
Figure \ref{K2_vs_r1_K1_0p03_vary_p2_1} reports the transition scenarios whenever $p_2$ is varied from $-0.05$ to $1$. For $p_2=0.1$ and $1$, the system is experiencing a continuous transition to synchronization. We observed that, similar to Fig.\ref{K2_vs_r1_diff_p2_1}, here also, the level of synchronization increases due to an increase in the effective triadic coupling strength. Moreover, the system synchronizes earlier in high $p_2$. On the other hand, for $p_2=-0.012$, solutions of the derived self-consistent equations show that a new $r_1$ curve with low $r_1$ value pops up in the $K_2-r_1$ space, depicted in Fig.\ref{K2_vs_r1_K1_0p03_vary_p2_1}(d). Unlike the previous case ($K_1=0.01$), this new $r_1$ curve separated the synchronization diagram into two portions with a decrease in $p_2$. Numerically simulated data points plotted on top of the analytical curves reveal the consistency between them. The numerical data points show that 
at $p_2=-0.012$ the system follows tiered synchronization transition for the adaptive initial condition process. Along with that, the system will follow the new $r_1$ curve with proper choice of the initial condition. 
At $p_2=-0.02$, the system jumps from strong synchronization state to a comparatively low synchronization state in backward simulation. At this point, the forward continuation of the simulation shows decrease in the synchronization level following the analytical curve. With more decrease in the $p_2$ value, the synchronization level of the lower curve decreases. These findings demonstrate the impact of $p_2$ and adaptation of the polynomial function to the fraction of nodes on the synchronization transitions with the variation of $K_2$.

One natural question arises, what will be the effect of this coefficient $p_2$ whenever $K_1$ is varied? To find this, we have plotted $r_1$ as a function of $K_1$ in Fig.\ref{K1_vs_r1_adapt_K2_polynomial} under the adaptation function $g_2(r_1)$. For $p_2=0$ the system follows double explosive transition when the nodes having degree $\geq 95$ are adapted \cite{dutta2025double}. An increase in the $p_2$ value pulls the backward transition point toward a lower coupling strength and leads the system to show an explosive synchronization transition. In contrast, the negative values of $p_2$ reduce the effect of $K_2$ and the transition becomes continuous. Figure \ref{K1_vs_r1_adapt_K2_polynomial} depicted the consistency between analytical and numerical results.     

\begin{figure}
    \centering
    \includegraphics[width=1\linewidth]{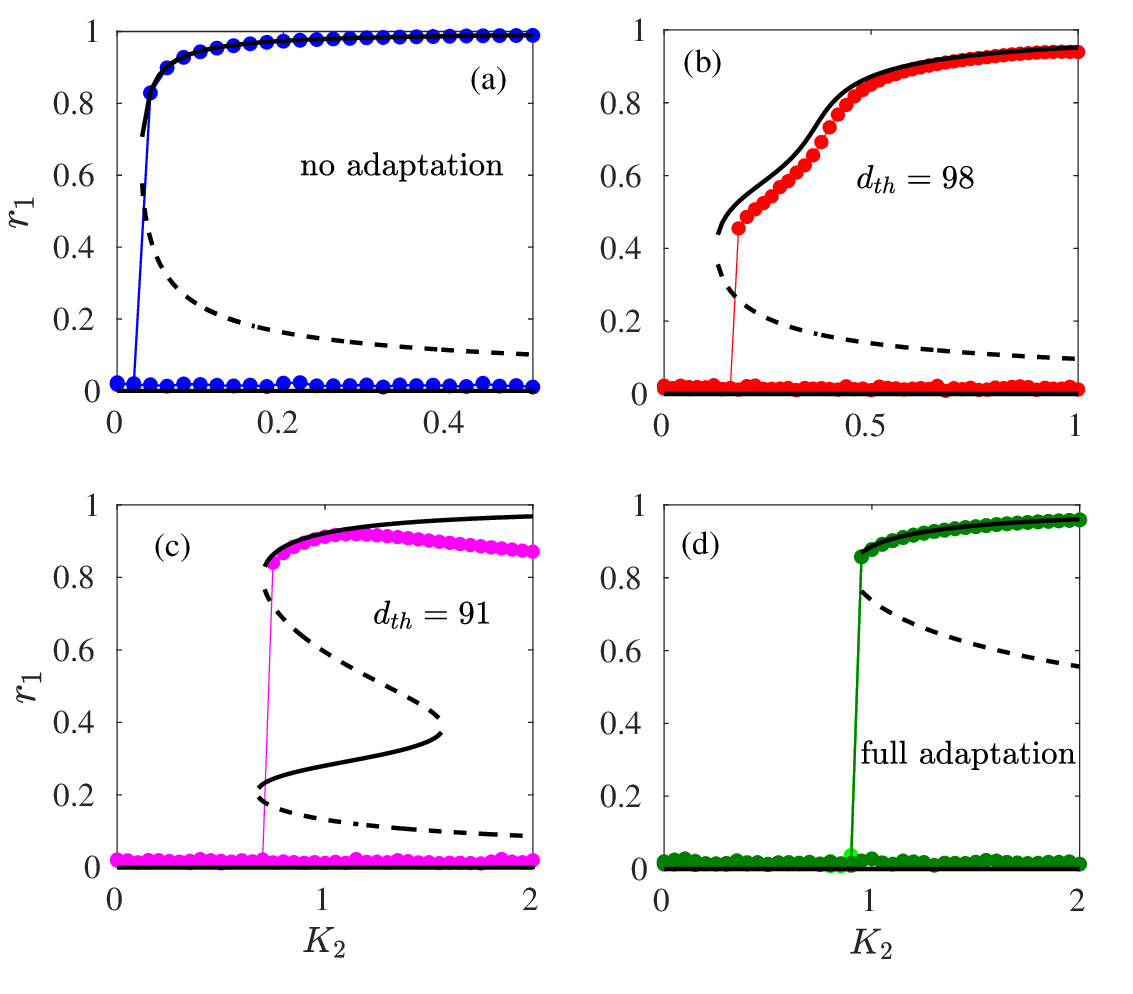}
    \caption{$r_1$ as a function of triadic coupling $K_2$ under under the adaptation function $g_{2}=A e^{-B(r_1-C)^2}$. The order parameter adaptation with the triadic coupling is given partially to (a) none of the nodes, (b) nodes having degree greater equal to $98$, (c) nodes having degree greater equal to $90$ and (d) all nodes. Other parameter values are taken as $K_1=0.01,~A=1,~B=1$ and $C=2.6$.}
    \label{K2_vs_r1_adapt_gaussian}
\end{figure}

Therefore, partial adaptation of power law functions of order parameter with the triadic coupling gives rise to a different type of explosive transition path (Fig.\ref{K2_vs_r1_diff_p2_1}(c)). In order to investigate the root behind this type of transition, first we check the dependence of the transition on the adaptation functions. We have considered a Gaussian function, $g_{2}(r_1)=A e^{-B(r_1-C)^2}$, whose characteristics are totally different from the power law function \cite{biswas2024effect}. We have done similar analysis of the self-consistent equations along with the numerical simulations. We have found that the partial adaptation of this function with the triadic coupling shows similar behavior as of the polynomial function (Fig.\ref{fig1}). The parameter values are kept fixed at $K_1=0.01,~A=1,~B=1$ and $C=2.6$. The gradual formation of explosive path with high level of synchronization from another explosive path with comparatively low level of synchronization is clearly demonstrated in Fig.\ref{K2_vs_r1_adapt_gaussian}. 

Therefore, from this analysis, we can conclude that this type of synchronization behavior is independent of adaptation function and depends only on the partial adapting scheme with the triadic coupling.

\section{Conclusion}
In summary, here we have shown the synchronization behavior of a networked system under partial order parameter adaptation scheme in the form of different functions like polynomial, Gaussian, etc. We have theocratically investigated the synchronization profiles by deriving a pair of self-consistent equations by using the Ott-Antonsen ansatz. The analytical findings have been validated by numerical simulation by considering a random network. Interestingly, here we have identified nontrivial synchronization diagrams under partial adaptation, i.e the functions of order parameter are adapted to a fraction of nodes of the system. Unlike the results reported in the Ref.\cite{dutta2025double}, which shows double explosive transition both in the forward and backward directions with the variation of the pairwise coupling $K_1$ due to partial adaptation of $r^b$ with the triadic coupling, here also, we found that for low $K_1$, the partial adaptation of $r^b$ induces double jumps in the explosive transition path by generating one stable and one unstable state, with variation of higher-order coupling strength. In the evaluation process of the transition paths from no adaptation to full adaptation, the critical point of the lower portion of the non zero order parameter branch moves faster toward higher coupling value than the upper portion, which generates a classical explosive path to another classical explosive path. However, for high $K_1$, this adaptation process induces a single jump in the transition path that takes the form of tiered transition.
\par In addition, we investigated the transition scenarios when both the coefficients of the polynomial adaptation function are non-zero. We observed that the positive values of the coefficients increase the synchronization level. Whereas, negative values of one coefficient of the adaptation function induce totally different behavior. Remarkably, the system shows a multistability between strong and weak synchronization state, where the synchronization level of the weak synchronization state decreases with increases in $K_2$. This happens due to the combined effect of positive and negative coefficients. Therefore, the interplay between the considered parameters induces several types of nontrivial transitions in the random network under the polynomial adaptation function. Furthermore, we have proved the generality of two-step synchronization behavior by taking the adaptation function as Gaussian. These results illustrate that the double jumps in the transition paths stem from the partial order parameter adaptation technique with the triadic coupling in a single-layer network configuration.      

\section{APPENDIX}
\begin{figure}[h!]
    \centering
    \includegraphics[width=1\linewidth]{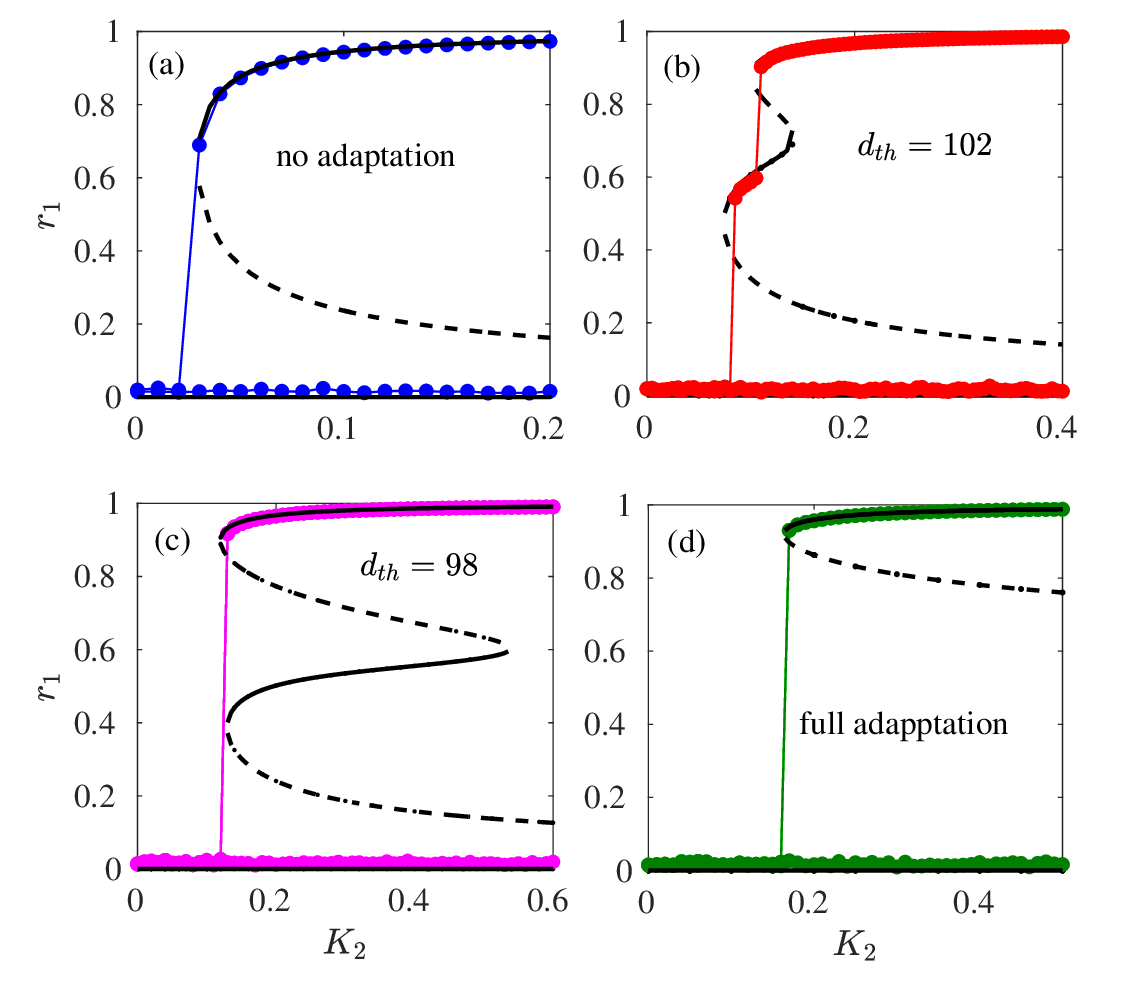}
    \caption{$r_1$ as a function of triadic coupling $K_2$ under the adaptation function $g_{2}=p_1r_1^{b}+p_2$. The order parameter adaptation with the triadic coupling is given partially to (a) none of the nodes, (b) nodes having degree greater equal to $102$, (c) nodes having degree greater equal to $98$ and (d) all nodes. Other parameter values are taken as $K_1=0.01,~m_1=1,~m_2=0,~p_1=1,~p_2=0,~a=0$ and $b=10$.}
    \label{fig1}
\end{figure}
We have adapted the order parameter $r_1$ to a fraction of nodes and observed the changes by increasing this fraction. 
From Fig.\ref{fig1}(a), we observed that when none of the nodes of the system are adapted, the $r_1$ curve clearly characterizes the typical explosive transition path. When we adapt a fraction of nodes, which have degree $\geq 102$, the $r_1$ curve folds once, generating two saddle-node bifurcation points. As a consequence, one stable and one unstable state appears along with the existing ones. From Fig.\ref{fig1}(b), it seems that the explosive path splits into two portions.
As soon as we increase the fraction of adapted nodes, this fold becomes deeper (Fig.\ref{fig1}(c)) and the backward transition points or the saddle node bifurcation points move on toward higher coupling value. Also the lower part of the non zero $r_1$ branch moves forward faster than the upper part. Finally, when all nodes are adapted, this lower part of the path vanishes and contains only the upper part with high $r_1$ value. We notice that the transition path with or without adaptation is explosive with different vertical widths between the stable and unstable parts because of low pairwise coupling.
\begin{figure}
    \centering
    \includegraphics[width=1\linewidth]{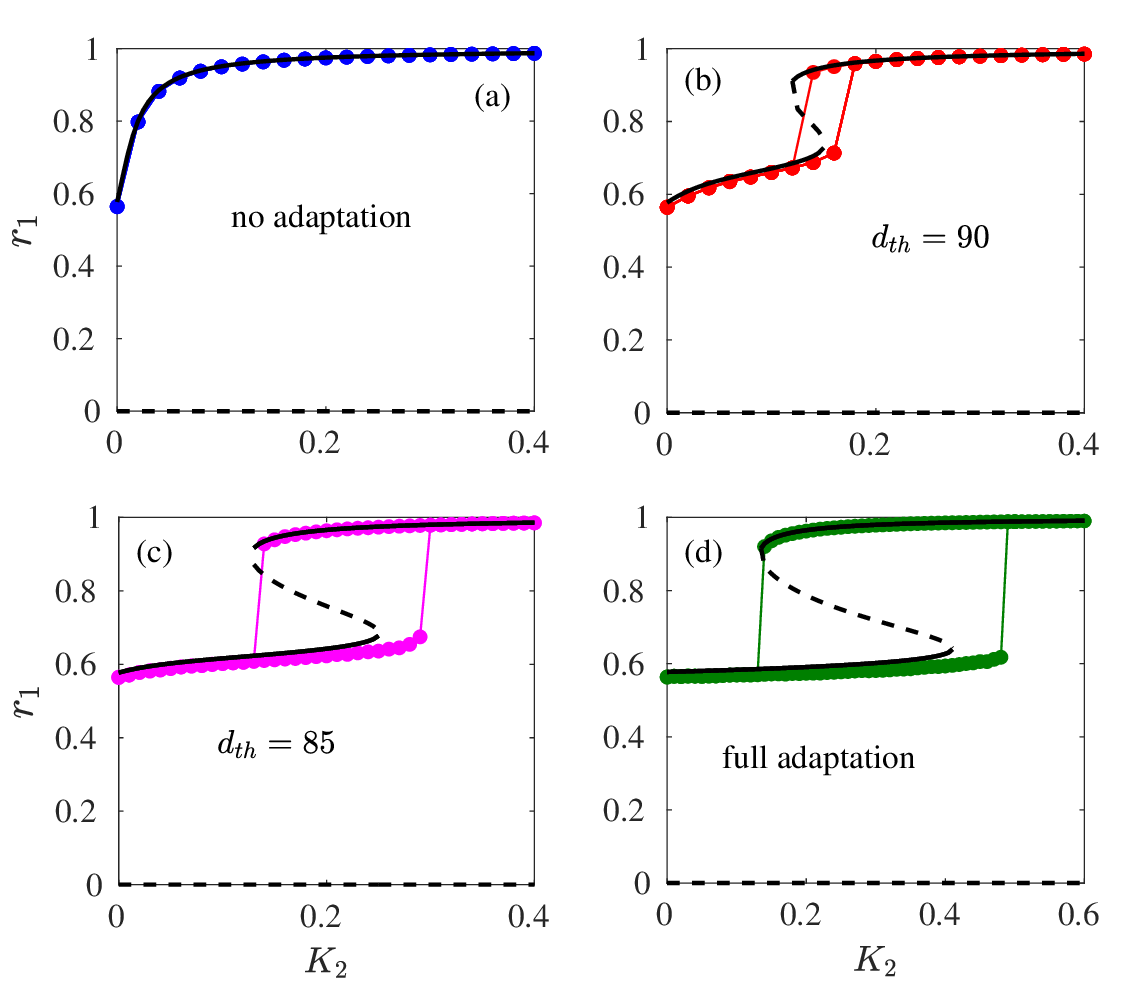}
    \caption{$r_1$ as a function of triadic coupling $K_2$ under the adaptation function $g_{2}=p_1r_1^{a,b}+p_2$. The order parameter adaptation with the triadic coupling is given partially to (a) none of the nodes, (b) nodes having degree greater than or equal to $90$, (c) nodes having degree greater than or equal to $85$ and (d) all nodes. Other parameter values are taken as $K_1=0.03,~m_1=1,~m_2=0,~p_1=1,~p_2=0,~a=0$ and $b=10$.}
    \label{fig2}
\end{figure}

In Fig.\ref{fig2} we have increased the $K_1$ value to $0.03$ and seen that it leads the non-adapted system to follow a continuous transition path (Fig.\ref{fig2}(a)). After a certain number of nodes are partially adapted, the continuous path is folded by generating an unstable part. This leads to a tiered synchronization transition (Fig.\ref{fig2}(b)). Moreover, the forward transition point moves to a higher coupling value with an increase in the number of adapted nodes. Also, the transition remains tiered for full adaptation. Here also we have put the numerically simulated order parameter values on top of the analytical curves, showing good agreement. Thus, Fig.\ \ref{fig1} and Fig.\ref{fig2} demonstrate the formation of explosive and tiered paths from explosive and continuous paths, respectively. 
\section{Acknowledgment}
S.D. acknowledges the support from DST, India under the
INSPIRE program (Code No. IF190605).
\bibliography{References}

\end{document}